\documentclass{article}

\usepackage{arxiv}

\usepackage[utf8]{inputenc} % allow utf-8 input
\usepackage[T1]{fontenc}    % use 8-bit T1 fonts
\usepackage{hyperref}       % hyperlinks
\usepackage{url}            % simple URL typesetting
\usepackage{booktabs}       % professional-quality tables
\usepackage{amsfonts}       % blackboard math symbols
\usepackage{nicefrac}       % compact symbols for 1/2, etc.
\usepackage{microtype}      % microtypography
\usepackage{lipsum}
\usepackage{graphicx}
\graphicspath{ {./images/} }

\title{An Architectural Design Decision Model for Resilient IoT Application}

\author{
 Cristovao Freitas Iglesias Jr \\
School of Electrical Engineering and Computer Science\\
Faculty of Engineering\\
University of Ottawa\\
  \texttt{cfrei096@uottawa.ca} \\
   \And
Claudio Miceli\\
PESC\\
Federal University of Rio de Janeiro\\
  \texttt{cmicelifarias@gmail.com} \\
  \And
Miodrag Bolic \\
School of Electrical Engineering and Computer Science\\
Faculty of Engineering\\
University of Ottawa\\
  \texttt{Miodrag.Bolic@uottawa.ca} \\
}

\begin{document}
\maketitle
\begin{abstract}
The Internet of Things is a paradigm that refers to the ubiquitous presence around us of physical objects equipped with sensing, networking, and processing capabilities that allow them to cooperate with their environment to reach common goals. However, any threat affecting the availability of IoT applications can be crucial financially and for the safety of the physical integrity of users. This feature calls for IoT applications that remain operational and efficiently handle possible threats. However, designing an IoT application that can handle threats is challenging for stakeholders due to the high susceptibility to threats of IoT applications and the lack of modeling mechanisms that contemplate resilience as a first-class representation. In this paper, an architectural Design Decision Model for Resilient IoT applications is presented to reduce the difficulty of stakeholders in designing resilient IoT applications. Our approach is illustrated and demonstrates the value through the modeling of a case.
\end{abstract}

% keywords can be removed

\section{Introduction}

Internet of Things (IoT) application is defined as a collection of automated procedures and data integrated with heterogeneous entities (hardware, software, and personnel) that interact with each other and with their environment to reach common goals \cite{1}. IoT application has caused a significant social and economic impact due to the application domain such as industrial, smart city, and healthcare domain \cite{2, 3}. Thus, any threat affecting the availability of IoT applications can be crucial financially and for the safety of the physical integrity of users. In this sense, one of the most critical domains is healthcare, where failures in monitoring patients' vital signs can significantly impact patient safety \cite{4}. This feature calls for IoT applications that remain operational and efficiently handle threats that could occur \cite{5}. However, designing an IoT application that can handle threats is a challenge for stakeholders due to the highly susceptible threats of IoT applications and the lack of a modeling approach that contemplates resilience as a first-class representation. In the following, will be discussed these difficulties.

High susceptibility to Threats of IoT applications occurs for several reasons. 
First, it is the fundamental characteristics of IoT that naturally predispose applications to potential failures. These include heterogeneity, interconnectivity, and expansive scale regarding devices, which collectively create a broad attack surface and multiple failure points. With the increasing sophistication of systems coupled with rising interoperability and maintenance issues, the complexity of managing such diverse objects continues to grow, surpassing the capabilities of human oversight \cite{6}.
Second, the typical deployment environments of IoT devices are often dynamic, uncontrolled, sometimes remote, and potentially hostile. The reliability of connections in such settings is generally low, providing ample opportunities for attackers to execute physical attacks and compounding the security management challenge \cite{6}.
Third, wireless technology for communication in IoT devices is another point of vulnerability. This mode of communication is inherently prone to interference and interception. This susceptibility makes it an easy target for adversaries who, with enough determination, could launch disruptive attacks such as Denial of Service.
Fourth, most IoT setup components, particularly end devices, need higher computing resources. This deficiency impedes the implementation of advanced security protocols, leaving the IoT components critically vulnerable to threats. The range of threats are diverse, encompassing communication loss between devices, process crashes, system unavailability due to power outages, malicious software, hacking attempts, inadequate security policies, physical accidents, malfunctions, outdated systems or software, and man-in-the-middle (MITM) attacks.
Finally, the dynamic nature of the IoT device's world, marked by rapid and unexpected context changes, contrasts starkly with the more stable environment of computers. Despite these changes, the expectation is for reliable IoT system functioning. To achieve robust and trustworthy IoT systems, incorporating redundancy at several levels and the ability to adapt automatically to changing conditions is essential \cite{7}. In essence, mitigating the high susceptibility of IoT applications to threats requires comprehensive strategies to enhance resilience, a critical focus of this study.
    
One way to minimize the abovementioned problems is to design an IoT application as a resilient system. A resilient system can resist various types of disturbances and recover fully or partially \cite{5,8}. One IoT application containing the constraints for resilience, such as redundancy, self-configure, self-heal, self-optimize, and self-protect, is a solution for dealing with any threat that may occur \cite{10}. However, resilience should be addressed in the early stages of the design phase, and one way of doing this would be with Architecture design decisions (ADDs) \cite{5,10}. They are considered first-class entities when architecting software systems. ADDs capture potential alternative solutions and the rationale for deciding among competing solutions \cite{34}. However, most of the models of ADDs present in the literature are generic. They do not present the necessary and specific concepts for dealing with the resilience design in IoT applications \cite{40}. The \cite{5} indicates that creating resilience meta-models for constructing models of resilient IoT applications can help the stakeholders in the design phase to create an architectural foundation. These models will be used to analyze all possible behaviors, reducing complexity and allowing a global view of the system due to the high level of abstraction. Once behaviors are recognized, understood, and classified in a model, they will be used as insights into architecting, designing, and engineering resilient ultra-large-scale systems.

Given the points raised above, we propose \textbf{A}rchitectural \textbf{D}esign \textbf{D}ecision for \textbf{R}esilient \textbf{IoT} (ADD4RIOT) Application. More specifically, it is a meta-model for designing a resilient IoT Application. It provides a common lexicon and taxonomy, defining the main resilient concepts and their relationships to model IoT applications able to handle threats, restore operations and adapt to environmental changes. It can enable a common understanding between stakeholders about a target resilient IoT system by providing an approach that helps to capture precisely state requirements and domain knowledge. ADD4RIOT can generate a primary representation of an IoT Application architecture from the point of view of resilience so that group of stakeholders can communicate. 

This paper brings four contributions: 
\begin{itemize}
    \item First, the requirements for modeling a Resilient IoT application are presented.
    \item Second, it defines Resilient IoT Applications based on the resilience requirements raised.
    \item Third, a Meta-Model to define a common understanding of a field of interest from the point of view of resilience, through the definition of its vocabulary and key resilient constraints.
    \item Fourth, it presents a modeling process to use the ADD4RIOT to design resilient IoT applications; such a process allows the separation of responsibilities between the different experts involved in constructing an IoT application.
\end{itemize}

This paper is organized as follows. Section 2 presents the requirements for modeling a resilient IoT application. In section 3, the ADD4RIOT is described. Section 4 presents the modeling process for resilient IoT application design. Section 5 illustrates the use and demonstrates value of the ADD4RIOT by modeling a case. We briefly discuss related work in section 6, and the paper concludes with future work and conclusions in section  7.

\section{Requirements for modeling a resilient IoT application}

Resilient systems can endure and successfully recover from disturbances by identifying problems and mobilizing the available resources to cope with the disturbance. Resiliency techniques allow a system to recover from disruptions, variations, and degradation of expected working conditions \cite{11,12}. A Biological system such as the Immune System (IS) is resilient \cite{12,13}. The Immune System is highly adaptive and scalable, able to cope with multiple data sources, fuse information together and make decisions. The IS has multiple interacting agents, operates in a distributed manner over multiple scales and has a memory structure to enable learning. The IS is considered an excellent example of a resilient system because it is a complex system that is in operation in most living beings on our planet and has been improving itself over millions of years through the process of evolution called natural selection \cite{14,5,13, 15}. Furthermore, the IS has already inspired some works in computer science \cite{12, 16,17}.

Given the advantages mentioned above of the IS, this paper inspires the resilience requirements for IoT application modeling on some fundamental resilience properties of the IS. The IS has five resilience key properties: (i) Monitoring, (ii) Detection, (iii) Protection, (iv) Restoration, and (v) Memorization \cite{15}. The following subsections describe these five key resiliency properties of the Immune System that are used as requirements for the resilient IoT application modeling.

\subsection{Monitoring}

\textbf{IS property: }The Immune System has cells called Leukocytes that are produced or stored in many locations in the body, including the thymus, spleen, and bone marrow. The two basic types of leukocytes are (i) phagocytes, cells that chew up invading organisms, and (ii) lymphocytes, cells that allow the body to remember and recognize previous invaders and help the body destroy them. The leukocytes circulate via lymphatic vessels and blood vessels between the organs and nodes. In this way, the immune system works coordinated to monitor the body for detecting germs or substances that might cause problems \cite{13, 15}.

\textbf{IoT requirement:} The monitoring of the resilient IoT application is essential. The monitoring should inspect operational resources, data flow, devices, services, and energy efficiency. The data from monitoring should be stored in Knowledge Base for other components such as protection, detection, and restoration to retrieve and perform their respective operations. Using an IoT Gateway could be an effective way to monitor the behavior of any system in several layers: application, network, and physical \cite{18, 19}. IoT gateways may not be used only for communication, to connect the sensors to the internet, or to collect data from sensors. Such IoT gateways can perform the function of monitoring the system. For example, if a sensor is damaged, it must be replaced automatically. An IoT gateway is essential to building an efficient, secure, and easy-to-maintain system \cite{20}.

\subsection{Detection}

\textbf{IS Property:} The immune system operates like an advanced detection system for pathogens (disease-causing agents such as bacteria and viruses). It identifies these harmful invaders by recognizing their unique molecular structures. Once identified, the immune system designs highly specific molecules, known as antibodies, which fit the pathogen's molecules with lock-and-key precision. These antibodies are not just identifiers; they are equipped with the mechanisms necessary to neutralize or destroy the pathogen, thus effectively eliminating the threat to the body's health \cite{13, 15}.

\textbf{IoT Requirement:} An essential first step for an IoT application in dealing with threats is their identification. Threats, defined as any potential danger to IoT systems, can arise from faults, failures, and errors, with the primary sources being natural events, hardware limitations, and human actions.
Natural threats encompass events like earthquakes, hurricanes, floods, fires, and power outages. These threats are uncontrollable and unpredictable, often causing severe disruptions to IoT systems.
Hardware threats stem from the physical and technical limitations of the IoT devices themselves. These can include energy and memory constraints, natural wear and tear, malfunctions, computational limitations, mobility issues, scalability concerns, communications media faults, device multiplicity, and low battery life.
Human threats, on the other hand, are actions by individuals or groups, either internal (those with authorized access) or external (entities operating outside the network), aiming to harm or disrupt an IoT application. These threats can target any of the three main layers of an IoT application: the Application Layer, Network Layer, and Physical Layer.
Given the diversity and complexity of these threats, the process of detection should begin with systematic monitoring to pinpoint the causes of threats. The data obtained from this detection process should be stored in a knowledge base, serving as a valuable resource for future threat restoration and application optimization. Depending on the specific scenario, various detection algorithms may be employed. The knowledge gained from identifying and understanding these threats can then update the knowledge base, equipping the IoT application with improved defenses for future protection.

\subsection{Protection}

\textbf{IS Property:} The immune system safeguards the body against external threats like bacteria, viruses, fungi, or parasites and internal threats like cancer cells. This protective property is carried out through redundant cells and a series of bodily reactions responding to infections or cancer cells \cite{13, 15}.

\textbf{IoT Requirement:} In an IoT application, protection serves dual roles: it acts both defensively and offensively against major threats. The defensive aspect focuses on shielding the system from threats and continuously updating the knowledge base with the new threat information. Conversely, the offensive aspect ensures the system can fend off recurrent attacks that have previously caused harm. Two key mechanisms employed to achieve protection are self-protection and redundancy, both of which can also aid in restoration.
Self-protection refers to safeguarding the entire system from threats. A comprehensive system failure can be a result of either a malicious attack or a cascading series of component failures. Self-protection strategies aim to counteract not only individual failures or attacks that might affect the entire system's behavior but also anticipate and prevent such situations \cite{livro-iota}. The importance of self-protection in the context of IoT is straightforward and hardly requires further emphasis.
Redundancy is a fundamental strategy for ensuring resilience. Typically, redundancy implies having more components within a system than is strictly required for functionality. Functional redundancy involves the overlap of functions across different components. It enables the substitution of a failed component with a different one, thus restoring all or part of the lost functionality \cite{10}.

\subsection{Restoration}

\textbf{IS Property:} The immune system not only identifies and eliminates pathogens but also ensures the body continues to function with minimal resources until a stable state is restored. This restoration is achieved through complex and organized processes, such as inflammation and healing \cite{13, 15}.

\textbf{IoT Requirement:} Restoration in an IoT application implies recovering the system to its normal functioning state after a catastrophic event. This component should administer healing to the weakened parts of the application, enabling them to resume their regular functions. Threat detection is accomplished by the detection component, aided by the monitoring component. Several strategies can be employed for restoration, including Redundancy, Self-Configuration, Self-Healing, Fault-Recovery, and Disaster Recovery.
Dynamic reconfiguration of components should be considered when workload increases to ensure optimal application performance. If full restoration is not achieved, Self-Optimization should be employed. Now let's describe some architectural constraints:
i) \textit{Self-Configuration:} This allows the system to readjust itself when the environment changes or when trying to achieve a set objective for the system \cite{livro-iota}.
ii) \textit{Self-Optimization:} This feature enables the system to gauge its current performance and compare it to an optimal performance level. The system will adjust its operations to approach optimal performance. It can also alter its operations to accommodate new user-set policies \cite{livro-iota}.
iii) \textit{Self-Healing:} This allows the system to recover from or avoid faults. Self-healing can be implemented in two different modes: reactive and proactive. In reactive mode, the system detects and recovers from faults as they occur and attempts to repair faulty functions if possible. In proactive mode, the system monitors its state to detect and adjust its behavior before reaching an undesired state \cite{livro-iota}.

\subsection{Memorization}

\textbf{IS Property:} The immune system's ability to rapidly and effectively respond to previously encountered pathogens is referred to as memorization. This response is possible due to the existence of a clonally expanded population of antigen-specific lymphocytes, which are immune cells that have learned to recognize and react to these specific threats \cite{13, 15}.

\textbf{IoT Requirement:} In a resilient IoT application context, the knowledge base component serves as its memory. This database maintains various information, including monitoring data, restoration data, a list of system vulnerabilities, and comprehensive data about different types of attacks and the corresponding preventive measures.
Monitoring data is gathered by the gateway, which cooperates with all system components and is essential for operating these components. Detection data collected during the detection phase contains information about the identified faults. The restoration component utilizes this data to facilitate the recovery of the resilient IoT application.

\section{ADD4RIOT}

Before delving into the details of the ADD4RIOT, we must first establish our working definition of a Resilient IoT Application. This is necessary due to the absence of a universally accepted definition, despite keen interest from both industry and academia.

We define a Resilient IoT Application as follows:

\textit{A set of interconnected infrastructures comprising connected devices, which facilitate their management, enable data extraction, and provide access to the data they generate, aiming to achieve common goals. These applications possess the ability to: 1) consistently monitor the system, 2) detect both new and existing threats that could harm the system, 3) protect the application from internal and external threats, 4) recover to a stable state and/or adapt its structure to function with minimal resources, and 5) record all impacts that threats may inflict on IoT Services, Resources, and Devices. This, in turn, facilitates faster and more effective responses to future threats.}

The ADD4RIOT model was developed to align with this definition and to aid in the design of Resilient IoT Applications. It aims to streamline the identification of common threats during the design process and encapsulate the core elements of resilience, i.e., monitoring, protection, recovery, and memory, as primary components of the architecture. It also proposes decision-making principles to guide the selection or rejection of countermeasures against potential threats.

ADD4RIOT consists of four main components: Inputs, Issues, Countermeasures (Mitigation strategies), and Decisions. The Inputs represent the elements of an IoT application's domain model, including critical objects that may be affected by threats. The Issues encapsulate the potential threats and call for possible Countermeasures. The Countermeasures component provides solutions that could mitigate these threats. Lastly, the Decisions component encapsulates the principles to guide the selection or rejection of countermeasures. These components collectively provide a comprehensive structure to facilitate the design of Resilient IoT Applications. In the following sections, we will elaborate on each of these components.

\begin{figure}[h]
  \centering
  \includegraphics[scale=0.5]{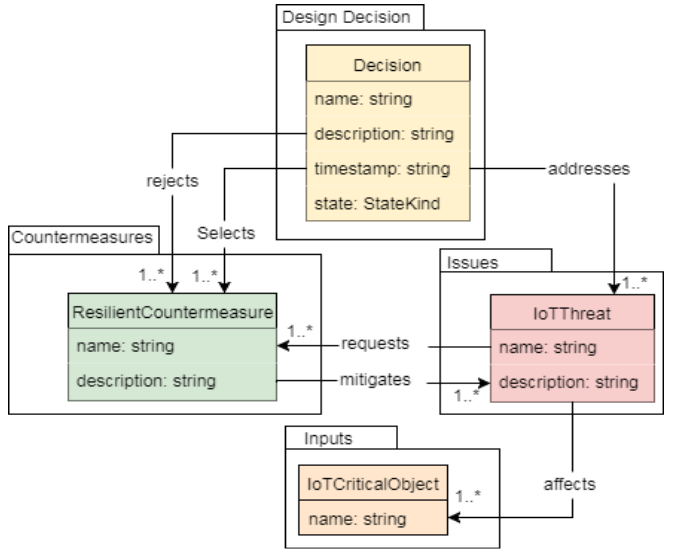}
  \caption{The four packages that composed the ADD4RIOT.}
  \label{packages}
\end{figure}

In summary, the ADD4RIOT brings i) the main threats described in the literature to speed up the threats identification procedure in the design process; ii) the tactics, constraints, and properties of resilience represented as the first class to make explicit and allow to capture potential alternative resilient solutions, iii) architectural design decisions principles, such as Issues (IoT Threats), Solutions (Resilient Countermeasures) and Decisions,  to support the decision of select or reject resilient countermeasures to mitigate the threats, and iv) Group decision Making principles to driving the way stakeholders make collaborative decisions. The ADD4RIOT is divided into four packages to facilitate the understanding and use of the meta-model. The four packages are Inputs (colorful elements in dark orange), Issues (colorful elements in red), Countermeasures (colorful elements in green), and Decisions (colorful elements in yellow). Figure \ref{packages} depicts all packages, the principal elements, and the relationships between them. The Inputs package contains classes representing the elements of an IoT application domain model, such as IoT Critical Objects that are affected by IoT Threats. The Issues package formalizes the concept of IoT Threats and requests Resilient Countermeasures as a possible solution. The Countermeasures package describes the concept of Resilient Countermeasures that mitigates IoT Threats. Lastly, the Decisions package formalizes the decision to select and reject a resilient countermeasure for addressing IoT Threats. The following will explain in detail each package.

\subsection{Input Package}

\begin{figure}[h]
  \centering
  \includegraphics[scale=0.7]{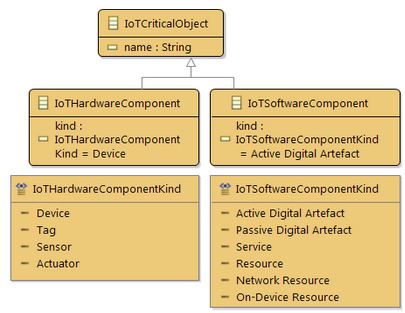}
  \caption{Input Package.}
  \label{input_package}
\end{figure}

The Inputs package is described in Figure \ref{input_package}. This package defines all ADD4RIOT inputs as IoT critical objects. It represents an element that can be affected by IoT threat and was proposed in \cite{24}. These IoT critical objects can be identified in an IoT application domain model, user story, storyline, and functional requirements. Examples of domain models are IoT Domain Model (IoT-DM) and Personalized Monitoring System Domain Model (PMS-DM) \cite{livro-iota, meupaper}. Through the Input package, we must identify the objects with the highest chance of being affected by a threat and damage to the system's functioning. A Critical Object can be IoT Hardware and  IoT Software components \cite{livro-iota}. The IoT Hardware components are classified as Device, Tag, Sensor, and Actuator. The  IoT software components are classified as: Active Digital Artefacts, Passive Digital Artefact, Service, Resource, Network Resource and On-Device Resource. Details about these components can be found in \cite{livro-iota,meupaper}.

% \clearpage
% \newpage

\subsection{Issue Package}

\begin{figure}[h]
\centering
\includegraphics[scale=0.5]{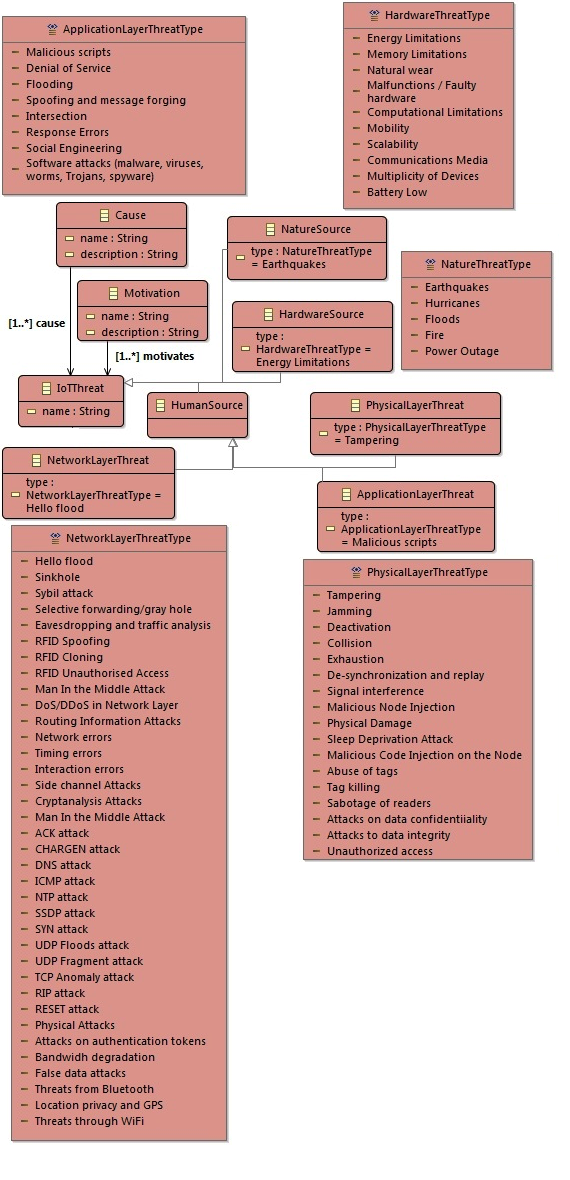}
\caption{Issues Package.}
\label{issues_package}
\end{figure}

The Issues package is highlighted in Figure \ref{issues_package} and has 15 elements. This package brings a set of concepts, entities, and relationships describing the main IoT threats that can damage an IoT application.
In ADD4RIOT, an IoT Threat is an action that takes advantage of security weaknesses in an IoT Application and has a negative impact on it. The Motivation and Cause elements should describe the IoT Threat. A Motivation element should explain why the IoT Threat is a problem, and the Cause element should explain the reason for this IoT Threat. IoT Threats can originate from three primary sources: Nature source, Hardware source, and Human sources. In \cite{pagina}, a detailed description of each threat type that composes the enumerations: Hardware Threat Type, Nature Threat Type, Application Layer Threat Type, Network Layer Threat Type, and Physical Layer Threat Type.

\clearpage
\newpage

\subsection{Countermeasures Package}

\begin{figure}[h]
  \centering
  \includegraphics[scale=0.4]{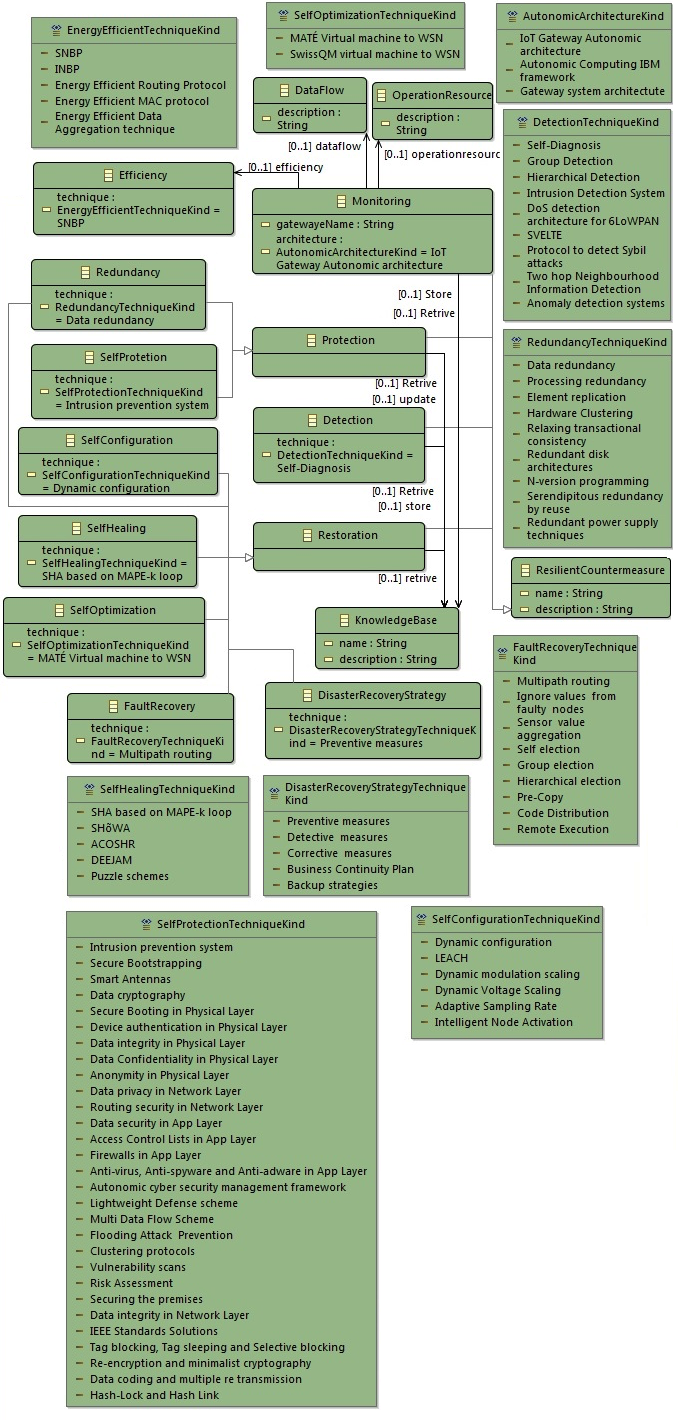}
  \caption[Short caption]{Countermeasures Package.}
    \label{countermeasures_package}
\end{figure}

This Package is highlighted in Figure \ref{countermeasures_package} and has 24 elements. It brings together a set of technologies available to implement the resilience properties that enable an IoT application to handle an IoT Threat. This Package has five fundamental properties that allow addressing the definition of Resilient IoT Application that are: Monitoring, Protections, Detection, Restoration and Memorization (knowledge base).

A Resilient Countermeasure can be classified in four properties: Monitoring, Protections, Detection and Restoration and interacts with knowledge base.

\textbf{Monitoring:}  it  performs  monitoring  on  Operational resources, Data flow, Energy Efficiency components and  help to protection, detection and restoration to work. The Monitoring of IoT application can be performed by a Gateway. The monitoring data will be stored in Knowledge base for other resilient solution to retrieve and perform the necessitated operations. The monitoring of IoT Application through IoT gateways can be performed using Autonomic Architectures (the architectures that compose the enumeration called Autonomic Architectures kind enumeration are described in \cite{pagina}).

\textbf{Protection:} it can be implemented through 9 tactics of Redundancy and 29 Self-protection techniques. The Protection update knowledge base when, for example, get a new attacks and retrieve old situation from knowledge base. The tactics that compose the enumeration Redundancy Technique Kind and Self Protection Technique Kind are described in \cite{pagina}.
    
\textbf{Detection:} it detects  the  vulnerabilities  and  the  weak  points  of  the  Resilient IoT Application.  The processes of detection is performed by monitoring through the implementation of Detection Techniques (all techniques that compose the enumeration Detection Technique Kind are described in \cite{pagina}) to find some IoT Threat. Detection techniques are algorithms  and  the vulnerabilities  detected  will be stored in  the  vulnerability  list  in  knowledge base  for  future Protection and can be utilized  by Restoration resilient solution in order to perform  restoration and  optimization  of  Resilient IoT Application.  
    
\textbf{Restoration:} the main responsibility of it is to bring back the Resilient IoT Application to its normal state after catastrophic  situation.  The restoration will  perform  Self-Healing  on  weakened parts of the Resilient IoT Application and empower them to perform their regular functions. The weak parts will be  detected  by  the  detection  resilient solution  with  the  help  of  the  monitoring  resilient solution.  Self-Configuration  will  be  used  to  reconfigure  the  components  when  workload  is  increased  to achieve optimization of Resilient IoT Application. Self-Optimization will also be used in case when full restoration is not achieved  by  the  restoration.  The Restoration resilient solution can implement Disaster Recovery Strategy like backup and contingency plans that are the best approaches to secure systems against natural threats. Finally, the restoration resilient solution can implement too Fault Recovery techniques important to deal with IoT Threats in WSN. All techniques to implement Restoration that compose the enumeration: Self Configuration Technique Kind, Self-Healing Technique Kind, Self-Optimization Technique Kind, Fault Recovery Technique Kind, Disaster Recovery Strategy Technique Kind are described in \cite{pagina}.

\clearpage
\newpage

\subsection{Decision Package}

The Decision Package is highlighted in Figure \ref{packages}. This Package combines Architecture design decisions and Group decision making principles and methods to enable group of stakeholders to find the best resilient countermeasure to be to address an IoT Threat identified by the Issues Package. This Package allows a model instantiated from ADD4RIOT be a primary representation of the architecture. A primary representation of architecture consists of architectural decisions and good architecture results from making good architectural decisions \cite{33}. Decision package has 13 elements and the main classes of this package is Decision, from is possible select or reject a resilient countermeasure to addresses an IoT Threat. The other elements of Decision Package use the same concepts of the two meta-models in the literature the Archium meta-model \cite{34} and Architecture design decisions Meta-model with Group Decision making \cite{35}. Due to limited space they are explain in details in \cite{pagina}.

% \begin{figure}[h]
%   \centering
%   \includegraphics[width=\linewidth]{decision.png}
%   \caption{Decision Package.}
%     \label{decision_package}
% \end{figure}

\section{Modelling Process}

\begin{figure}[h]
  \centering
  \includegraphics[scale=0.6]{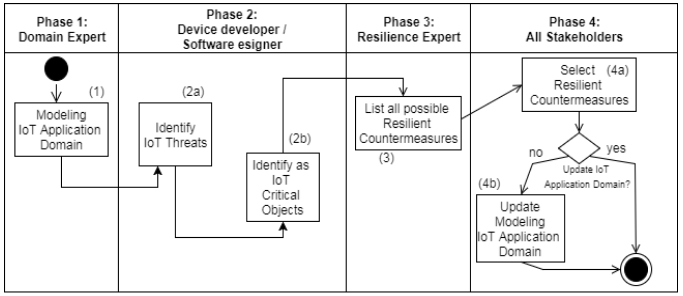}
  \caption{UML Activity Diagram describing the ADD4RIOT modelling process.}
      \label{activities}
\end{figure}

This section presents the steps required to design a resilient IoT application using the ADD4RIOT. It is divided into four phases that are executed in an iterative way as depicted in Figure \ref{activities}. The activities are performed by the main group of IoT application stakeholders. The set of actors in our process is composed of:

\textbf{Domain expert:} responsible for instantiates the IoT Domain model by identifying the domain elements, such as virtual entities, resources, devices, services and users. It has ability to understand domain concepts, including the data types produced by the sensors, consumed by actuators, accessed from storages, user interactions, and how the system is divided into regions.

\textbf{Resilience Expert:} responsible for identifying the critical objects in IoT Domain Model and lists the threats and associated countermeasures. It has experience on fault, failures and error in IoT device and software, and knowledge on security, self-management and resilient constraints.

\textbf{Device developer:} responsible for writing drivers for the sensors, actuators, storages, and end-user applications used in the domain. It has a deep understanding of the inputs/outputs, and protocols of the individual devices.

\textbf{Software designer:} responsible for defining the structure of an IoT application by specifying the software components and their generate, consume, and command relationships. It has Software architecture concepts, including the proper use of interaction modes such as publish/subscribe, command, and request/response for use in the application.  

\textbf{Network Manager:} responsible for install the application on the system at hand; this process may involve the generation of binaries or bytecode, and configuring middleware. It has a deep understanding of the specific target area where the application is to be deployed.

\clearpage
\newpage

Figure \ref{activities} depicts the UML Activity Diagram illustrating the modelling process with its phases and actors (or stakeholders).

\textbf{The first phase} of the modelling process (Phase 1) encompasses the modelling IoT Application (activity 1a). This activity is performed by Domain Expert that should instantiate the application domain model, since from it will be identified the IoT Threats and IoT Critical Objects. For this activity could be used any Domain meta-model of literature such as IoT Domain Model (IoT-DM) \cite{livro-iota} and Personalized Monitoring System Domain Model (PMS-DM) \cite{meupaper}.

\textbf{The second phase} of the modelling process (Phase 2) encompasses two activities:  Identify IoT Threats (activity 2a) and IoT Critical Objects (activity 2b).  These activities will be performed by Device developer and Software designer and to help them carry out these activities we provide a table called \textit{Relationship between IoT Application Domains, IoT Critical Objects and IoT Threats} (available in \cite{pagina}) based on \cite{37,38,39}. It gathers the main IoT Threats and critical IoT Objects for the three main domain of IoT Application: Industrial, Smart city and Health well-being. In this table we have the relation between the main elements that can be classified as IoT Critical Objects and the main source of IoT Threats that can affect the workings of these elements. Furthermore, the \textit{Relationship between IoT Application Domains, IoT Critical Objects and IoT Threats} table has relation with \textit{IoT Threats} Tables available in \cite{pagina}.

\textbf{The third phase} of the modelling process (phase 3) encompasses one activity: List possible Resilient Countermeasures (activity 3). It will be performed by Resilience Expert that should list the main resilient countermeasures to mitigate the IoT threats identified in the previous phase. To help the Resilience Expert carry out this activity the enumeration tables of ADD4RIOT (available in \cite{pagina}) gathers the main Resilient Countermeasures. Furthermore, the \textit{IoT Threats Tables} has relation with \textit{Resilient Countermeasures tables}. This relation is of many to many.

\textbf{The fourth phase} encompasses two activities.  Select the Resilient Countermeasures (activity 4a). In this activity all stakeholders will participate and use architecture design decisions and group decision using principles and methods through Decision package. It will help to find the best resilient Countermeasures from of the selected in the previous phase, in activity (3). Others stakeholders can be included in the group to participate in modelling processes such as software architects, developers, designers, testers, users, etc. Update modelling IoT Application Domain Model (activity 4b). Before finished the modelling processes must be checked if there is need update in IoT Application due to selection made in activity 4a.   

\section{Case: Resilient Nursing Home IoT application}

In this section, we validate our approach by applying it on a case. It is used to illustrate the utilization of ADD4RIOT modelling process to generate a primary representation of a Resilient Nursing Home IoT Application architecture. First, the case is introduced, after the modelling processes of resilient IoT application case is presented in more detail.

\subsection{Case Overview}

The nursing home IoT application case was inspired in \cite{22}. Our case presents the design of an IoT application that aims to perform early detection, rapid, and appropriate response to help in the monitoring of patients who are under care in separate rooms into a nursing home. Therefore, this application also must be resilient because any failure of the system can cause serious damage to patients. The application includes sensors that are being used by patients to capture vital signs and alerts when signals outside the normal patterns are detected. For example, if occurs an abnormal increase in body temperature (that can indicate some infection), the medical staff receive one alert.

\subsection{Modelling of Resilient Nursing Home IoT Application}

\subsubsection{Phase 1: This phase encompasses the activity related to modelling of Nursing home IoT application domain. }

\begin{figure}[h]
  \centering
  \includegraphics[scale=0.65]{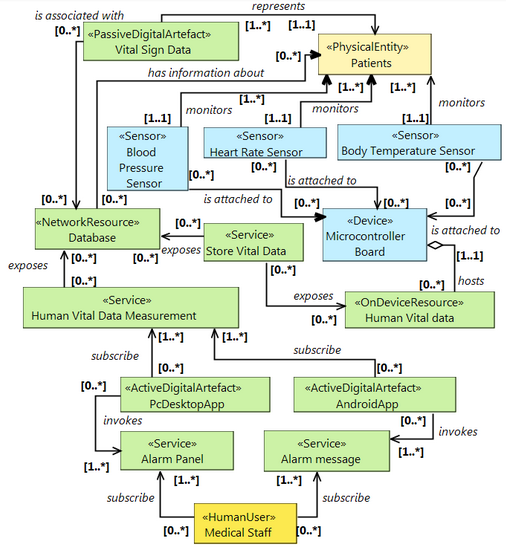}
  \caption{Nursing Home IoT Application Domain Model instantiated from IoT Domain Meta-Model \cite{livro-iota}.}
      \label{Nursing_Home}
\end{figure}

\textbf{Activity 1:} This modelling was done using the IoT Domain Meta-model of IoT-ARM \cite{livro-iota} and is depicted in Figure \ref{Nursing_Home}.
The element HumanUser represents the medical team that subscribes to the alarm service, provided by the system through an Android app or a desktop application. The Android and Desktop applications are Active Digital Artefacts (from the IoT Domain Meta-Model) and invoke their respective alarm services: Alarm Panel and Alarm message to alert the medical staff in case of abnormal situation with patients. Alarms represent the communication system interface with the user. The Alarm Panel service is invoked by the desktop computer application and displays an alarm on the PC screen. On the other hand, the Alarm message service is invoked by the Android application and displays a message on the mobile screen of the users that make up the medical team. The Android and Desktop applications to invoke the alarm services them subscribes another service called Human Vital Data Measurement that will read the user vital data from the Database and evaluate if they are out of normal ranges for human health. The database is represented by an element of type Network Resource and is associated with an element of type PassiveDigitalArtefact called Vital Sign Data that represent a Physical Entity called Patients. The vital data captured by sensors are inserted in the database through a service called Store Vital Data that receives the data from the resource Human Vital data, which is an element of type On-DeviceResource that is hosted within the Device. In this case, the On-DeviceResource is a software component that provides a way to connect to the data obtained by the sensors. For example, this data can be exposed through an XBee/ZigBee network. The Device is represented by a microcontroller board, for example, an Arduino, which is connected to the three types of sensors that are collecting the vital patient data. The three types of sensors are: Blood Pressure Sensor, Heart Rate Sensor, and Body Temperature Sensor.

% \clearpage
% \newpage

\subsubsection{Phase 2: This phase raises the issues that can damage the Nursing Home IoT application.}
\begin{figure}[h]
  \centering
  \includegraphics[scale=0.9]{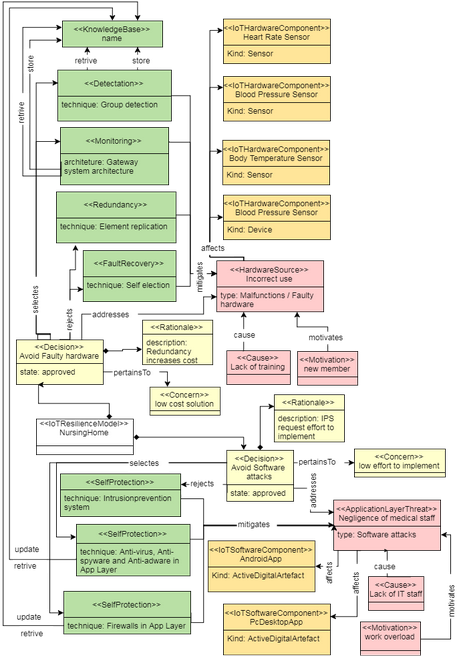}
  \caption{Resilient Nursing Home IoT Application Model instantiated from ADD4RIOT.}
    \label{Nursing_Home_res}
\end{figure}

\textbf{Activity 2a:} The Software attacks and Malfunction/Faulty hardware are the threats that will be addressed in Nursing Home IoT application. i) The Software attacks can occur due to Negligence of medical staff, because changes and updates in desktop computer configuration can cause system malfunctioning. Medical staff may install contaminated software upgrades that propagates virus into the desktop computer. A cause of this can be lack of IT team to take care of security and the work overload of medical staff. ii) The Malfunctions/Faulty hardware can occur due to incorrect use, because improper use for a long period can lead to a malfunction and can cause interruption of availability of application. A motivation of this is hiring a new member for medical team. A cause of this can be Lack of training. This IoT Threats are represented by the colored elements in red in the Figure \ref{Nursing_Home_res}.
\textbf{Activity 2b:} A total of six elements were classified as IoT Critical Objects  in function of IoT threats selected in previous activity and with the help of table
called \textit{Relationship between IoT Application Domains, IoT Critical Objects and IoT Threats} available in \cite{pagina}. The elements that were identified as IoT critical object are: i) Active Digital Artefacts called PcDesktopApp and AndroidApp and ii) Device called Microcontroller Board and the Sensors called Blood Pressure, Heart Rate and Body Temperature. These are the main objects that can be affected by the identified threats in the activity 2a. See colorful elements in orange on Figure \ref{Nursing_Home_res}.

\clearpage
\newpage

\clearpage
\newpage

\subsubsection{Phase 3: this phase raises the some possibles Resilient Countermeasures in functions of Threats and critical object identified in the Nursing Home IoT application domain model.}

\textbf{Activity3:} Selection made based on the enumeration tables available in \cite{pagina}, where the references and the explanation for each tactic can be found. The Figure \ref{Nursing_Home_res} exposes the countermeasures (see colorful elements in green). 

Four countermeasures were selected to mitigate the Malfunctions/Faulty hardware that the Incorrect use of sensors and devices can cause in application. 

\textbf{i) Monitoring using IoT Gateway Autonomic architecture.} In \cite{23} was presented an intelligent architecture which consists of a large number of sensing objects for monitoring purposes that can be used in IoT application. An embedded-based gateway for use in a monitoring system was proposed in an IoT network. The gateway is a critical component for collecting, recording and forwarding data obtained from sensors. It is programmable, low-cost, real-time and flexible. The software and hardware for wired and wireless communication interfaces are successful and suitable for field trials.

\textbf{ii) Detection using Group Detection.} The goal of fault detection is to verify that the services being provided are functioning properly, and in some cases to predict if they will continue to function properly in the near future. In \cite{24} a detection mechanism is proposed to identify faulty sensor nodes. Algorithm is based on the idea that sensors from the same region should have similar values unless a node is at the boundary of the event-region. The algorithm start by taking measurements of all neighbors of a node and uses the results to calculate the probability of the node being faulty

\textbf{iii) Redundancy using Element replication.} The structure and functionality of a replica are exactly the same as that element so that they can substitute each other without problem \cite{livro-iota}. In case of Nursing Home IoT application the sensors and device could be replicated.

\textbf{iv) Fault Recovery using Self-Election.} When passive replication is applied, the primary replica receives all requests and processes them. In order to maintain reliability  between  replicas,  the  state  of  the  primary  replica  and  the  request  information  are  transferred  to  the  backup replicas atraves de self-election \cite{25}.
   
Three countermeasures were selected for mitigate  Software attacks that Negligence of medical staff can allow that affect the software component of application.

\textbf{i) Self-Protection using Intrusion prevention system. }To follow this tactic the Cumulative-Sum-based Intrusion Prevention System (CSIPS)  can  be  applied. It which detects malicious behaviors, attacks and distributed attacks launched to remote clients and local hosts based on the Cumulative Sum (CUSUM) algorithm \cite{26}.  

\textbf{ii) Self-Protection using Anti-virus, Anti-spyware and anti-adware in application layer.} Security software like antivirus or anti spyware is important for the reliability, security, integrity  and confidentiality of the IoT system and desktop computer.  

\textbf{ii) Self-Protection using Firewalls in application layer. }This is an extra effective layer of security that will help block attacks that authentication, encryption and ACLs would fail to do so. Authentication and encryption passwords can be broken if weak passwords were selected. A firewall can filter packets as they are received, blocking unwanted packets, unfriendly login attempts, and DoS attacks before even authentication process begins.

All these countermeasures are related to the knowledge base. It is a resilience requirement that the application has memorization of all occurrences.

\subsubsection{Phase 4: Select the countermeasures that best fit the concerns of the Resilient Nursing Home IoT application stakeholders.
}

\textbf{Activity 4a: }Here the stakeholders of the case made two decisions: i) the Decision to Avoid Malfunctions/Faulty hardware was to select Detection using Group detection technique and Monitoring using Gateway system architecture as resilient countermeasures. Because to achieve the concern of low cost solution the rationale is reject Redundancy using Element replication and Fault Recovery using self-election. Since this increase in the number of sensors and devices and thus increases the cost of the project. ii) the Decision to Avoid software attacks was to select self-protection using Anti-virus, Anti-spyware and Anti-adware in application layer and self-protection using Firewalls in application layer as resilient countermeasures. Because to achieve the concern of low effort to implement this solutions in the application the rationale is reject Self-protection using Intrusion Prevention system. Since the implementation will require the use of complex algorithms that will order more qualified workers  in project.
\textbf{Activity 4b:} Update in Nursing Home IoT application Domain Model. As a gateway is going to be used it will be necessary to update the model by inserting this new element.

\section{Related Work}

Some projects try to address the challenge of deal of resilience in Iot Application. 
But them present some drawbacks. Here we will highlight three important projects developing IoT architectures \cite{27, 28}.  

IoT-A \cite{livro-iota} is a European project that proposes an Architectural Reference Model (ARM) for IoT. But the IoT ARM is not an IoT architecture per se, but a set of best practices, guide-lines, and a starting point to generate specific IoT architectures \cite{27}. The unique resiliency treatment presented by IoT-ARM is through an architectural perspective called Availability and Resilience in which it presents a design choice catalog with only 9 generic tactics used in software architecture. The ADD4RIOT presents a total of 82 tactics to implement the resilient constraints such as redundancy, self-configure, self-heal, self-optimise and self-protect specific for IoT application architecture.
BeTaaS \cite{29} is a project that proposes an architecture for the IoT and Machine-to-machine (M2M) communication, to enable running applications over a local cloud of gateways. Betaas focuses in Dependability, but presents aspect of resilience. It is handled via the Failure Analysis Approach component that is responsible for the identification of potential causes of failures and for providing solutions to properly manage them. However presents no concept of ADDs and GDM in order to select possible solutions for failures as well as ADD4RIOT features. In addition to not specifying which elements can and should be classified as critical.
The EU FP7 OpenIoT research project, has introduced an IoT architecture \cite{30}. OpenIoT is based on IoT-ARM to achieve alignment, architecture development and specification.  OpenIoT Address resilience only partly and places the focus on resilience in terms of mitigation. For that, OpenIoT maintains an up-to-date inventory of entities and dynamically restructures the dependencies between entities, e.g., reconnects a service to another sensor in case of sensor failure. Thus, fail-over and recovery are integral parts of OpenIoT.

The projects above mentioned are solutions to specific problems of system information and do not consider the specification of resilience for IoT application. Furthermore, they do not provide elements to express resilience accurately in the early stages of development. They do not provide modelling mechanisms that contemplate resilience as first class representation to design a resilient IoT application like ADD4RIOT.

\section{Conclusion}

In this paper, we presented an Architectural Design Decision Model for Resilient IoT Application, called ADD4RIOT, due to high susceptible to threats of IoT Application and lack of modelling approaches contemplating resilience as first class representation to design Resilient IoT Application.
It provides a common lexicon and taxonomy, defining the main resilient concepts and their relationships, and a modelling process needed to generate a common understanding and facilitate decision making between stakeholders about a target resilient IoT application in question.
The  ADD4RIOT  concepts were  exemplified with the modelling of a Resilient Nursing Home IoT Application.
The modelling of the case allow to note how ADD4RIOT can reduces the difficulty of design an IoT application with resilient concepts.
The ADD4RIOT generated a primary representation of Resilient Nursing Home IoT Application architecture so that group of stakeholders were able to communicate. The ADD4RIOT allowed to identify IoT Critical Objects in Nursing Home IoT Application domain and IoT Threats that could affect them. Next, it was possible to find possible Resilient Countermeasures in functions of IoT Threats and IoT Critical Objects, and select which would be the best ones for the case.
In  the  future  work, we  intend  to integrate ADD4RIOT in an framework to automatic design  resilient IoT applications.

\clearpage
\newpage
\bibliographystyle{unsrt}  
\bibliography{template}  %%% Remove comment to use the external .bib file (using bibtex).

% %%% Comment out this section when you \bibliography{references} is enabled.

% George Kour and Raid Saabne.
% \newblock Real-time segmentation of on-line handwritten arabic script.
% \newblock In {\em Frontiers in Handwriting Recognition (ICFHR), 2014 14th
%   International Conference on}, pages 417--422. IEEE, 2014.

% George Kour and Raid Saabne.
% \newblock Fast classification of handwritten on-line arabic characters.
% \newblock In {\em Soft Computing and Pattern Recognition (SoCPaR), 2014 6th
%   International Conference of}, pages 312--318. IEEE, 2014.

% \bibitem{hadash2018estimate}
% Guy Hadash, Einat Kermany, Boaz Carmeli, Ofer Lavi, George Kour, and Alon
% \newblock Estimate and replace: A novel approach to integrating deep neural
%   networks with existing applications.
% \newblock {\em arXiv preprint arXiv:1804.09028}, 2018.

% \end{thebibliography}

\end{document}